\documentclass[12pt,twoside]{article} 
\usepackage{graphics}
\usepackage{lscape}
\usepackage{multirow}
\usepackage{epsfig}
\usepackage{cite}
\usepackage{color}
\date{}

\begin{document}

\centerline{\bf Adv. Studies Theor. Phys., Vol. 7, 2013, no. 16, 759 - 771} 

\centerline{\bf HIKARI Ltd, \ www.m-hikari.com}

\centerline{\bf http://dx.doi.org/10.12988/astp.2013.3674} 

\centerline{} 
\centerline{}

\centerline {\Large{\bf Modeling the Dynamics of Infectious Diseases}} 

\centerline{} 

\centerline{\Large{\bf in Different Scale-Free Networks}} 

\centerline{} 

\centerline{\Large{\bf with the Same Degree Distribution }} 

\centerline{} 

\centerline{\bf {Raul Ossada, Jos\'{e} H. H. Grisi-Filho,}}
\centerline{\bf {Fernando Ferreira and Marcos Amaku}} 

\centerline{} 

\centerline{Faculdade de Medicina Veterin\'{a}ria e Zootecnia} 

\centerline{Universidade de S\~{a}o Paulo} 

\centerline{S\~{a}o Paulo, SP, 05508-270, Brazil} 


\bigskip

{\footnotesize Copyright $\copyright$ 2013 Raul Ossada et al. This is an open access article distributed under the Creative Commons Attribution License, which
permits unrestricted use, distribution, and reproduction in any medium, provided the original work is properly cited.}

		\begin{abstract}
		The transmission dynamics of some infectious diseases is related to the contact structure between
individuals in a network. We used five algorithms to generate contact networks 
with different topological structure but with the same scale-free degree distribution. 
We simulated the spread of acute and chronic infectious 
diseases on these networks, using SI (Susceptible -- Infected) 
and SIS (Susceptible -- Infected -- Susceptible) epidemic models. 
In the simulations, our objective was to observe the effects of the topological structure 
of the networks on the dynamics and prevalence of the simulated diseases. 
We found that the dynamics of spread of an infectious disease
on different networks with the same degree distribution may be considerably different. 
\end{abstract}


	\section{Introduction}
    \label{sec:introduction}	
		The contact pattern among individuals in a population is an essential factor for the spread of
infectious diseases. In deterministic models,
the transmission is usually modelled using a contact rate function, 
which depends on the contact pattern among individuals and also
on the probability of disease transmission. The contact function among individuals with different ages, 
for instance, may be modelled
using a contact matrix~\cite{AndersonMay} or a continuous function~\cite{Amaku}. However, 
using network analysis methods, we can investigate
more precisely the contact structure among individuals and analyze the effects of this structure
on the spread of a disease. 

The degree distribution $P(k)$ is the fraction of vertices
in the network with degree $k$. Scale-free networks
show a power-law degree distribution 
\begin{equation}
P(k) \sim k^{-\alpha} \, ,
\end{equation}

\noindent where $\alpha$ is a scaling parameter. 
Many real world networks~\cite{Caldarelli,Amaral,Clauset} are scale-free.
In particular, a power-law distribution of the number of sexual partners for 
females and males was observed in a network of human sexual contacts~\cite{Liljeros2001}.
This finding is consistent with the preferential-attachment mechanism
(`the rich get richer') in sexual-contact networks and, as mentioned
by Liljeros et al.~\cite{Liljeros2001}, may have epidemiological
implications, because epidemics propagate faster in scale-free networks
than in single-scale networks.

Epidemic models such as the Susceptible--Infected (SI) and 
Susceptible--Infected--Susceptible (SIS) models
have been used, for instance, to model the transmission dynamics of 
sexually transmitted diseases~\cite{Chen2006}
and vector-borne diseases~\cite{Shi2008}, respectively.
Many studies have been developed about the dissemination of diseases in 
scale-free networks~\cite{Lloyd18052001, PhysRevE.64.066112, PhysRevLett.86.3200, Zhou2006}
and in small-world and randomly mixing networks~\cite{Christley2005}.

Scale-free networks present a 
high degree of heterogeneity, with many vertices with a low number of contacts
and a few vertices with a high number of contacts. 
In networks of human contacts or animal movements, for example, 
this heterogeneity may influence the potential risk of spread of acute 
(e.g. influenza infections in human and animal networks, or foot-and-mouth disease in animal populations) 
and chronic (e.g. tuberculosis) diseases. Thus, simulating the spread of 
diseases on these networks may provide insights on how to prevent and control them.

In a previous publication~\cite{Grisi2013},
we found that networks with the same degree distribution may show 
very different structural properties. 
For example, networks generated by the Barab\'{a}si-Albert (BA) method~\cite{Barabasi15101999}
are more centralized and efficient than the networks generated by other methods~\cite{Grisi2013}.
In this work, we studied the impact of different structural properties on the dynamics
of epidemics in scale-free networks, 
where each vertex of the network represents an individual or even a set of individuals 
(for instance, human communities or animal herds). We developed routines to simulate 
the spread of acute (short infectious period) and chronic (long infectious period) 
infectious diseases to investigate the disease prevalence (proportion of infected vertices) 
levels and how fast these levels would be reached in networks with the 
same degree distribution but different topological structure,
using SI and SIS epidemic models.

		This paper is organized as follows. In Section~\ref{sec:hypothetical}, 
we describe the scale-free networks generated. In Section~\ref{sec:model}, 
we show how the simulations were carried out. 
The results of the simulations are analyzed in Section~\ref{sec:results}. 
Finally, in Section~\ref{sec:conclusions}, we discuss our findings. 
		
	\section{Scale-free Networks}
	\label{sec:hypothetical}
		We generated scale-free networks following the Barab\'{a}si-Albert (BA) 
algorithm~\cite{Barabasi15101999}, using the function barabasi.game($n$, $m$, directed) from the R package igraph~\cite{Team_2008, rpkg:igraph}, varying the number of vertices ($n$ = $10^{3}$, $10^{4}$ and $10^{5}$), the number of edges of each vertex ($m$ = 1, 2 and 3) and the parameter that defines if the network is directed or not (directed = TRUE or FALSE). For each combination of $n$ and $m$, 10 networks were generated. Then, in order to guarantee that all the generated networks would follow the same degree distribution and that the differences on the topological structure would derive from the way the vertices on the networks were assembled, we used the degree distribution from BA networks as input, to build the other networks following the Method A (MA)~\cite{Grisi2013}, Method B (MB)~\cite{Grisi2013}, Molloy-Reed (MR)~\cite{springerlink:10.1007/978-3-540-44485-5_1}
and Kalisky~\cite{springerlink:10.1007/978-3-540-44485-5_1} algorithms, all of which were implemented and described in detail in Ref.~\cite{Grisi2013}. 
As mentioned above, these different networks have distinct structural properties. 
In particular, the networks generated by MB are decentralized and with a larger number of components,
a smaller giant component size, and a low efficiency when compared to the
centralized and efficient BA networks that have all vertices in a single component. 
The other three models (MA, MB and Kalisky) generate networks with intermediate characteristics between MB and BA models.

	\section{Model of the Epidemic Spread Simulations}
	\label{sec:model}
		The element $ij$ of the adjacency matrix of the network, $A$, is defined as $a_{ij}=1$ if there is an edge between
vertices $i$ and $j$ and as $a_{ij}=0$, otherwise. 

	We also define the elements of the vector of infected vertices, $I$. If vertex $i$ is infected, 
then $I_{i}=1$, and, if it is not infected, $I_{i}=0$.

	The result of the multiplication of the vector of infected vertices, $I$, by the
adjacency matrix, $A$, is a vector, $V$, whose element $i$ corresponds to the number of   
infected vertices that are connected to the vertex $i$ and may transmit the infection
\begin{equation}
V = I \cdot A .
\end{equation}

	Using Matlab, the spread of the diseases with hypothetical parameters along the vertices of the network was simulated using the following algorithm:
		
				\begin{enumerate}
					\item A proportion ($\pi_{0}$) of the vertices is randomly chosen to begin the simulation infected.  For our simulations, $\pi_{0} = 50\%$, since we are interested in the equilibrium state and this proportion guarantees that the disease would not disappear due to the lack of infected vertices at the beginning of the simulations.
					\item In the SIS (Susceptible -- Infected -- Susceptible) epidemic model, a susceptible vertex 
can get infected, returning, after the infectious period, to the susceptible state. For each time step:
\begin{enumerate}	
\item We calculate the probability ($p_{i}$) of a susceptible vertex $i$, that is connected to $V_{i}$ infected vertices, to get infected, using the following equation:
\begin{equation}
p_{i} = 1 - { (1-\lambda)^{V_{i}} }, 
\end{equation}								
where $\lambda$ is the probability of disease spread.

\item \label{item2.b} We determine which susceptible vertices were infected in this time step:
if $\delta_{i}\sim$Uniform(0,1) $\leq p_{i}$, the susceptible vertex becomes infected.
For each vertex infected, we generate the time ($\gamma_{i}$) that the vertex will be infected following a uniform distribution:
$\gamma_{i}\sim$ Uniform ($t_{min}$, $t_{max}$), where $t_{min}$ and $t_{max}$ are, respectively, the minimum and the maximum time of the duration of the disease.
										
						\item Decrease in 1 time step the duration of the disease on the vertices that were already infected, verifying if any of them returned to the susceptible state;
						\item Update the status of all vertices.
					\end{enumerate}

			\end{enumerate}

			For the SI (Susceptible -- Infected) epidemic model, we chose $\gamma_{i}$ in order to 
guarantee that an infected vertex remains infected until the end of the simulation.
		
			Varying the values of the parameter $\gamma_{i}$, we simulated the behaviour of hypothetical acute and chronic diseases, using different values of $\lambda$, considering that once a vertex gets infected it would remain in this state during an average fixed time (an approach that can be used when we lack more accurate information about the duration of the disease in a population) or that there would be a variation in this period, representing more realistically the process of detection and treatment of individuals (Table~\ref{tab:table1} shows the diseases simulated and the values of $\gamma_{i}$ assumed).

		\begin{table*}[htbp]		
		\centering
			\begin{tabular}{|c|c|c|c|c|}
				\hline
				{\bf Spreading} & {\bf Hypothetical} & \multirow{2}{*}{\bf Time} & \multicolumn{2}{c|}{\bf $\gamma_{i}$}\\
				\cline{4-5}
				{\bf Models} & {\bf Diseases} &  & $t_{min}$ & $t_{max}$\\
				\hline
				SI & Chronic & Fixed & \multicolumn{2}{c|}{$T+100$}\\
				\hline
				\multirow{5}{*}{SIS} & \multirow{2}{*}{Acute} & Fixed & \multicolumn{2}{c|}{30}\\
				\cline{3-5}
				 &  & Variable & 15 & 45\\
				\cline{2-5}
				 & \multirow{3}{*}{Chronic} & Fixed & \multicolumn{2}{c|}{360}\\
				\cline{3-5}
				 &  & Variable 1 & 330 & 390\\
				\cline{3-5}
				 &  & Variable 2 & 180 & 540\\
				\hline

			\end{tabular}
			\caption{\small Values of the parameter $\gamma_{i}$, used in the spreading models, where $T$ is the total time of simulation. For the SI model, we chose $\gamma_{i}$ in order to guarantee that an infected vertex remains infected until the end of the simulation.}
			\label{tab:table1}
		\end{table*}

			We adopted a total time of simulation ($T$) of 1000 arbitrary time steps. 
		
			For each spreading model, we carried out 100 simulations for each network, calculating the prevalence of the simulated disease for each time step. Then we calculated the average of the prevalence of these simulations on each network. After that we grouped the simulations by network algorithm. Finally, we calculated the average prevalence of these network models.

	\begin{figure}[htbp]
	\centering
	\scalebox{0.57}{\includegraphics{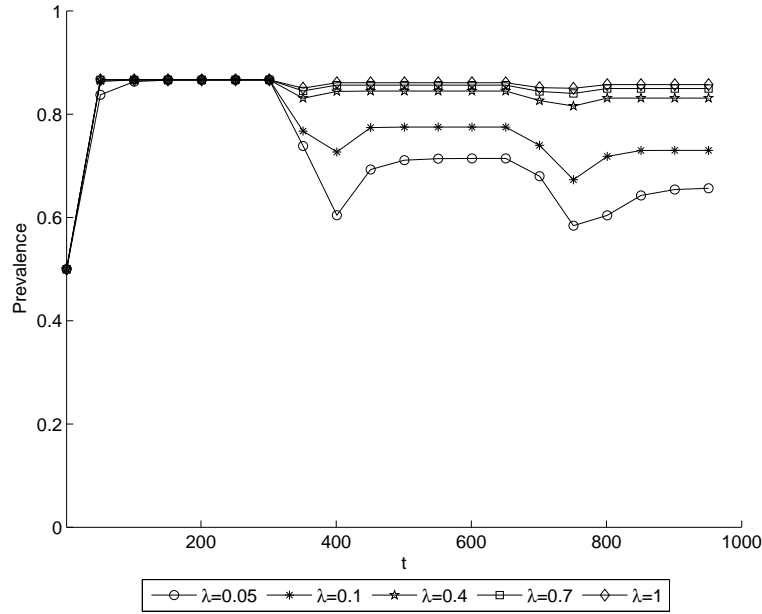}}
	\caption{\small 
	Effect of the increase of the probability of spreading ($\lambda$) on the prevalence in undirected networks with
	100000 vertices and $m=1$ generated by MB. 
	We assumed a SIS model for a chronic infectious disease with an infectious period
	ranging between 330 and 390 time units (variable 1).  }
	\label{fig:figure1}
	\end{figure}
		
		
	\section{Results}
	\label{sec:results}
		\subsection{Epidemic Spread Simulations on Undirected Networks}
			On the undirected networks, we observed that the disease spreads independently of the value of $\lambda$ used, 
and that an increase in $\lambda$ leads to an increase in the prevalence of the infection (Figure~\ref{fig:figure1}). 
Also, we observed that the prevalence tends to stabilize approximately 
in the same level despite the addition of vertices (figure not shown).

\begin{figure*}[htb]
	\centering
	\scalebox{0.54}{\includegraphics{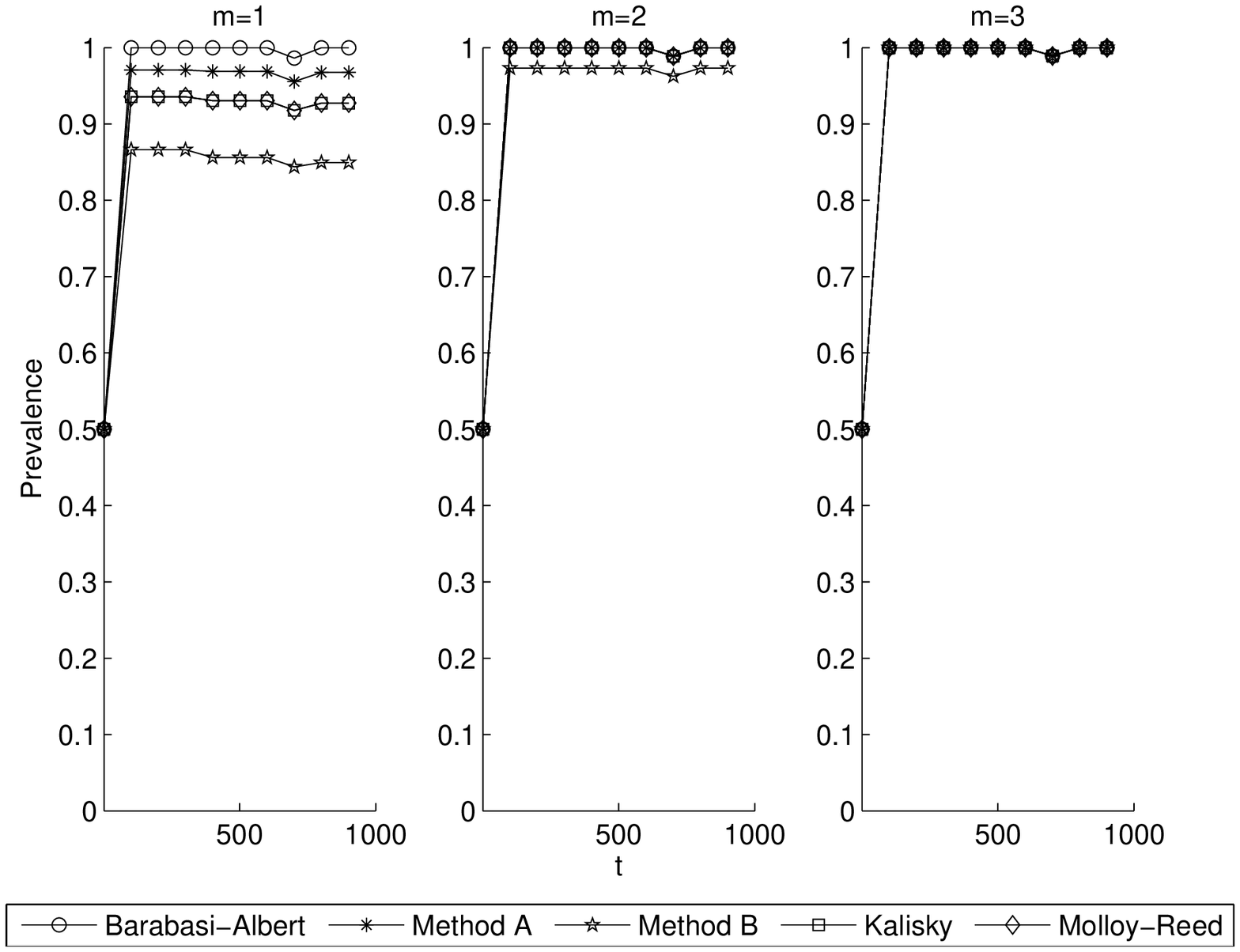}}
	\caption{\small
	Effect of the increase in the number of edges of each vertex ($m$) on the prevalence in undirected networks with
	100000 vertices and $\lambda=0.7$ generated by the methods indicated in the legend. We assumed a SIS model for a 
	chronic infectious disease with an infectious period ranging between 330 and 390 time units (variable 1). }
	\label{fig:figure3}
	\end{figure*}

			When we increase the number of edges of each vertex, there is an increase in the prevalence of the infection. A result that stands out is that, when $m = 1$, there is a great difference in the equilibrium level of the prevalence in each network. However, as we increase the value of $m$, the networks tend to show closer values of equilibrium (Figure~\ref{fig:figure3}).
		
\begin{figure*}[htbp]
	\centering
	\scalebox{0.6}{\epsfig{figure=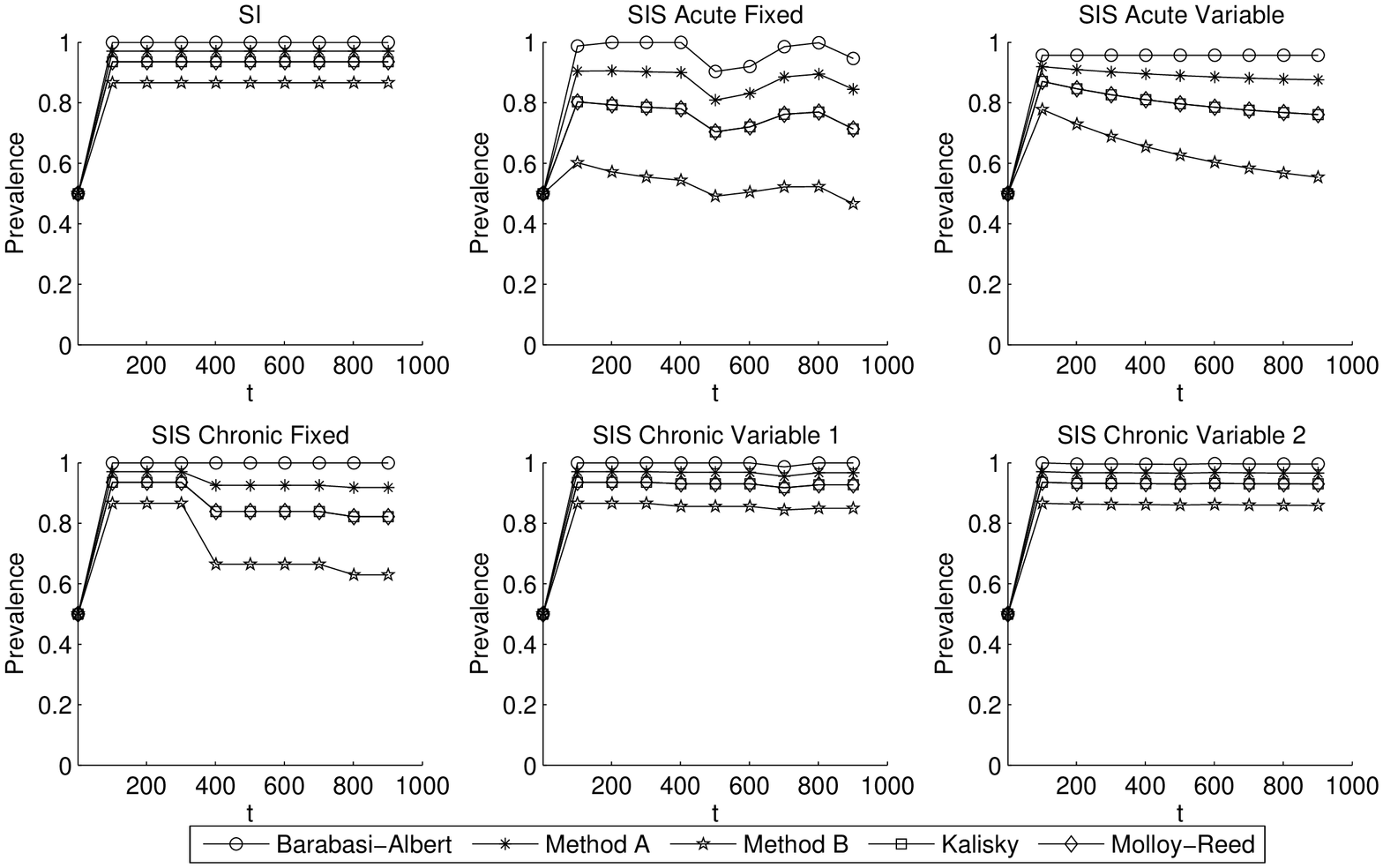, angle=-90}}
	\caption{\small 
	Results of the spreading simulations on the undirected networks with 100000 vertices, $m=1$ and $\lambda=0.7$, 
	considering SI and SIS models for acute and chronic infectious diseases. }
	\label{fig:figure4}
	\end{figure*}

	Among the undirected networks, the networks generated using the MB algorithm presented 
the lowest values of prevalence in the spreading simulations (Figure~\ref{fig:figure4}).

		\subsection{Epidemic Spread Simulations on Directed Networks}
\begin{figure*}[htbp]
	\centering
	\scalebox{0.57}{\includegraphics{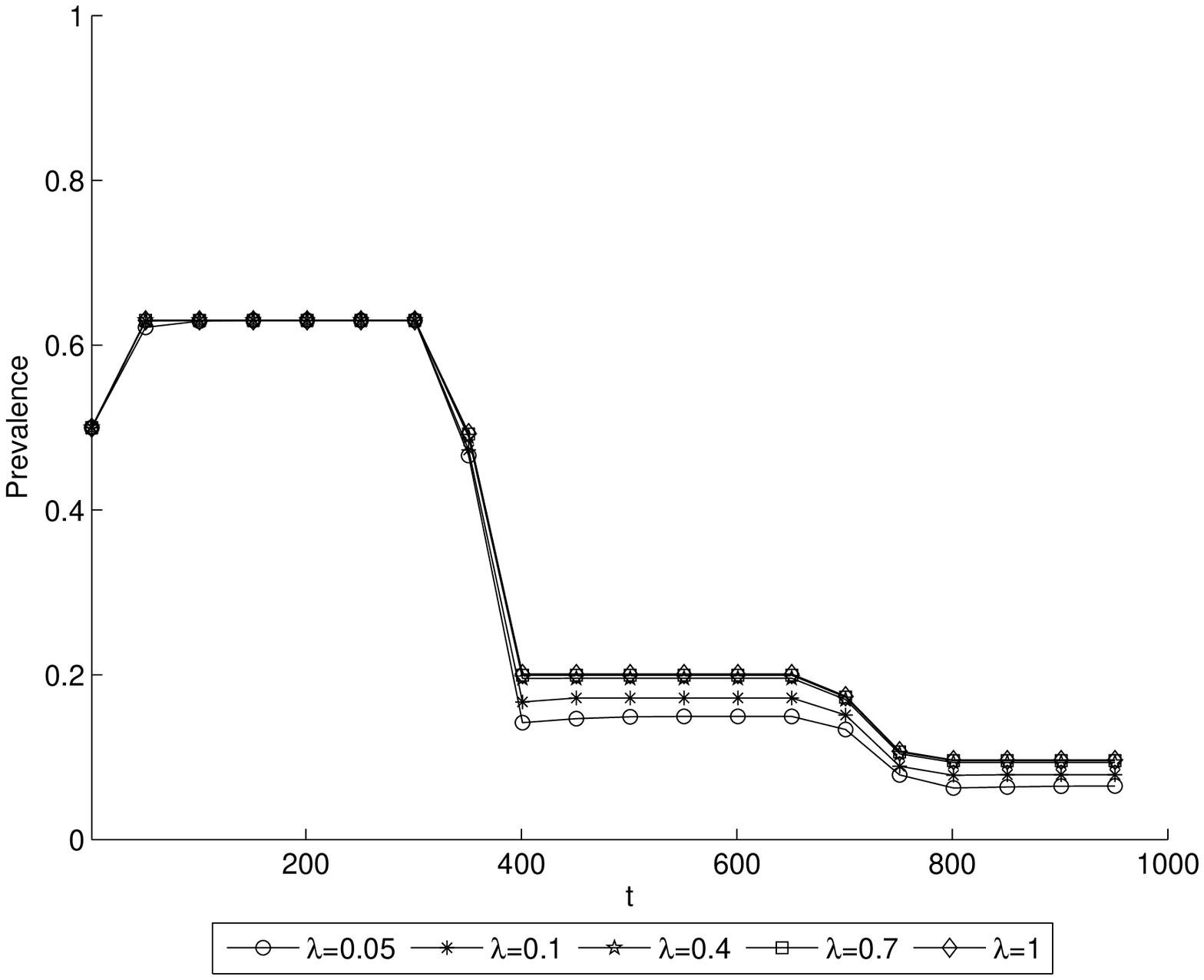}}
	\caption{\small
	Effect of the increase of the probability of spreading ($\lambda$) on the prevalence in directed networks with
100000 vertices and $m=1$ generated by MB. We assumed a SIS model for a chronic infectious disease 
with an infectious period
	ranging between 330 and 390 time units (variable 1). }
	\label{fig:figure5}
	\end{figure*}

			On the directed networks, we observed that, despite the simulations of acute diseases, the disease spreads independently of the value of $\lambda$ used and, as in the undirected networks, an increase in $\lambda$ leads to an increase in the prevalence of the infection (Figure~\ref{fig:figure5}). 
Also, similarly to what was observed for undirected networks,
the prevalence tends to stabilize approximately in the same level despite the addition of vertices (figure not shown).

\begin{figure*}[tbp]
	\centering
	\scalebox{0.55}{\includegraphics{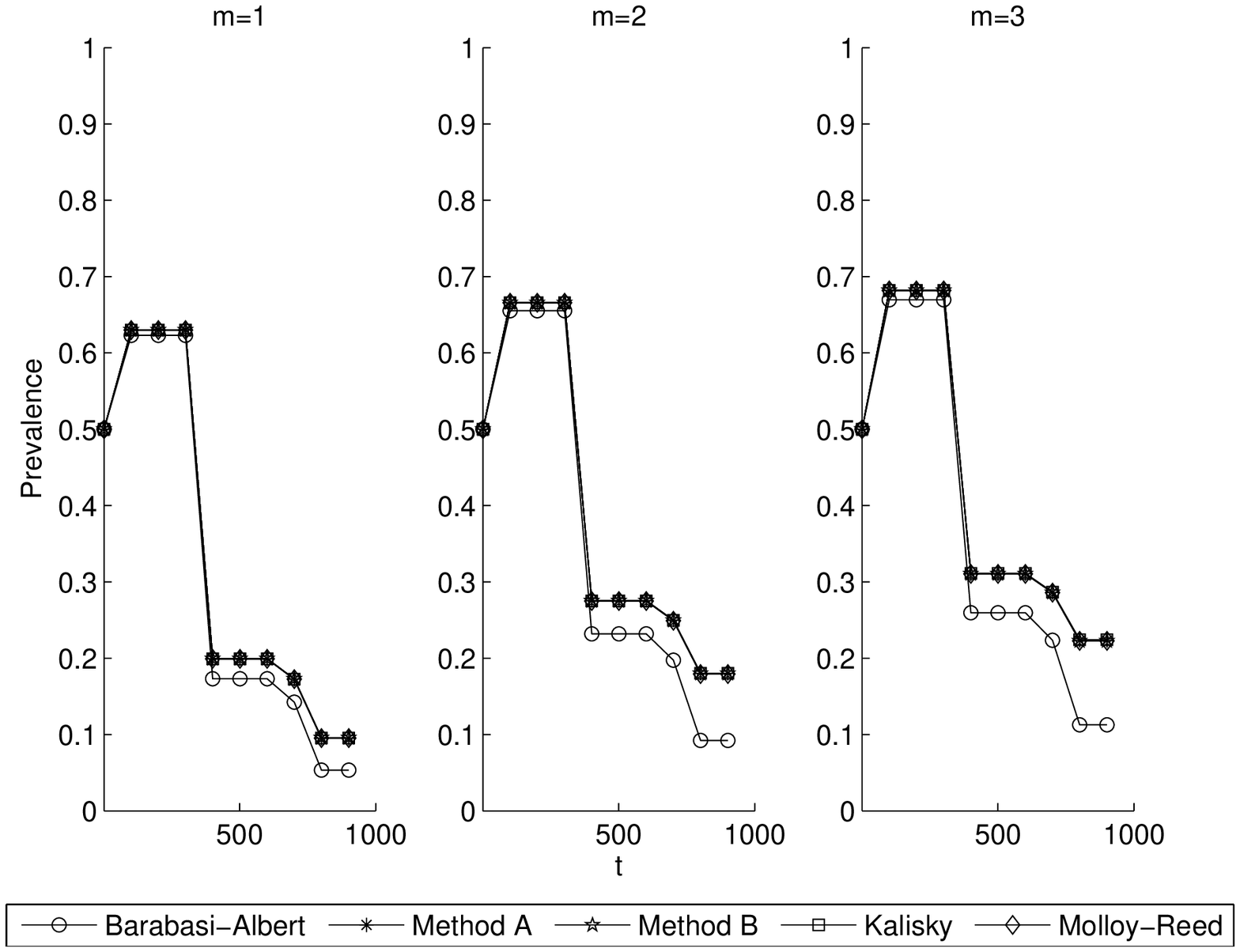}}
	\caption{\small
	Effect of the increase in the number of edges of each vertex ($m$) on the prevalence in directed networks with
	100000 vertices and $\lambda=0.7$ generated by the methods indicated in the legend. We assumed a SIS model for a 
	chronic infectious disease with an infectious period ranging between 330 and 390 time units (variable 1).}
	\label{fig:figure7}
	\end{figure*}

			When we increase the number of edges of each vertex, there is an increase in the prevalence of the infection (Figure~\ref{fig:figure7}). A result that stands out is that, when $m = 3$, the prevalence in the Kalisky networks tend to stabilize in a level a little bit higher than the other ones.
			
\begin{figure*}[htbp]
	\centering
	\scalebox{0.55}{\epsfig{figure=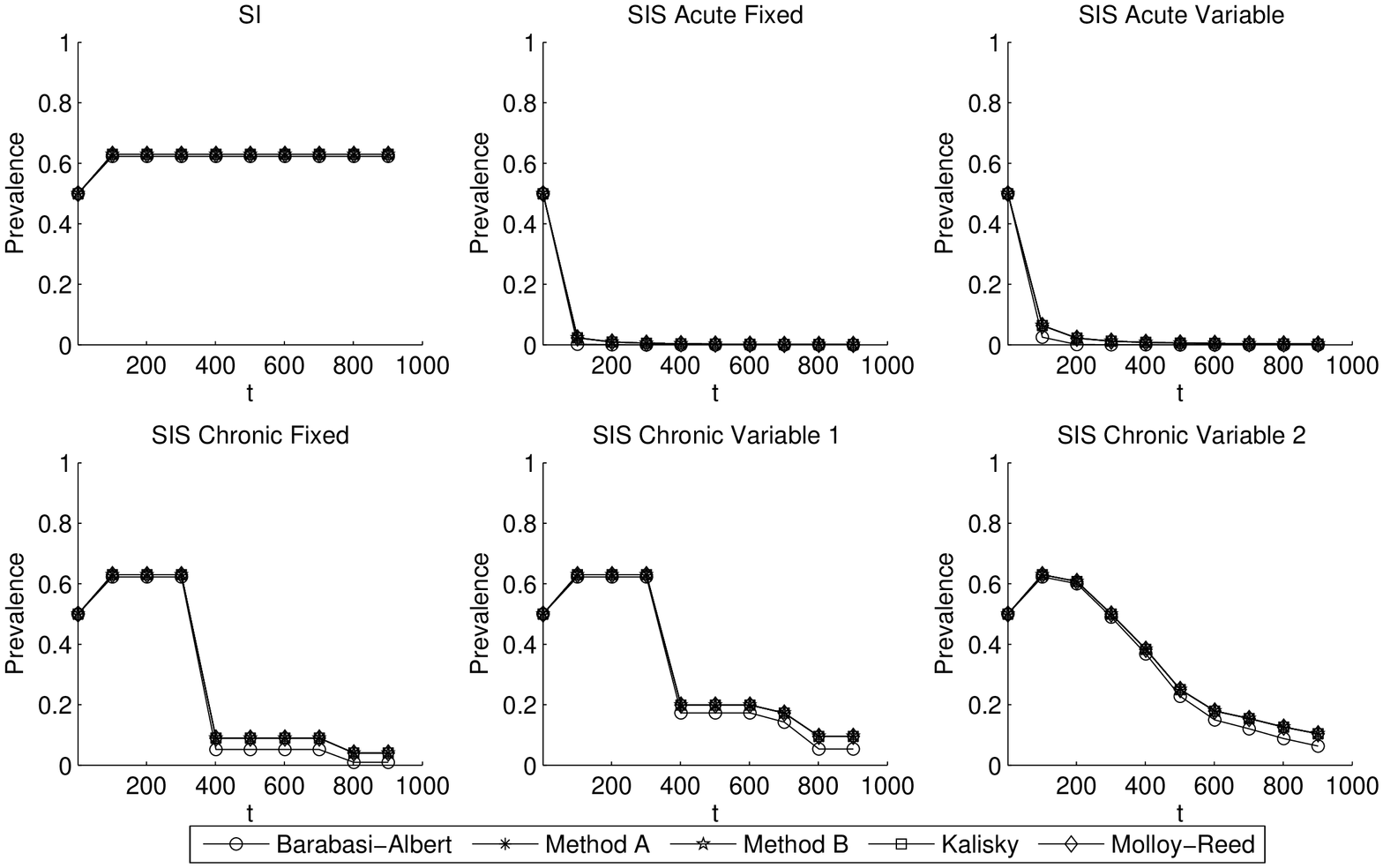, angle=-90}}
	\caption{\small
	Results of the spreading simulations on the directed networks with 100000 vertices, $m=1$ and $\lambda=0.7$, 
	considering SI and SIS models for acute and chronic infectious diseases.}
	\label{fig:figure8}
	\end{figure*}

			Among the networks, those generated using the BA algorithm presented the lowest values of prevalence in the spreading simulations (Figure~\ref{fig:figure8}).
			
\begin{figure*}[htbp]
	\centering
	\scalebox{0.58}{\includegraphics{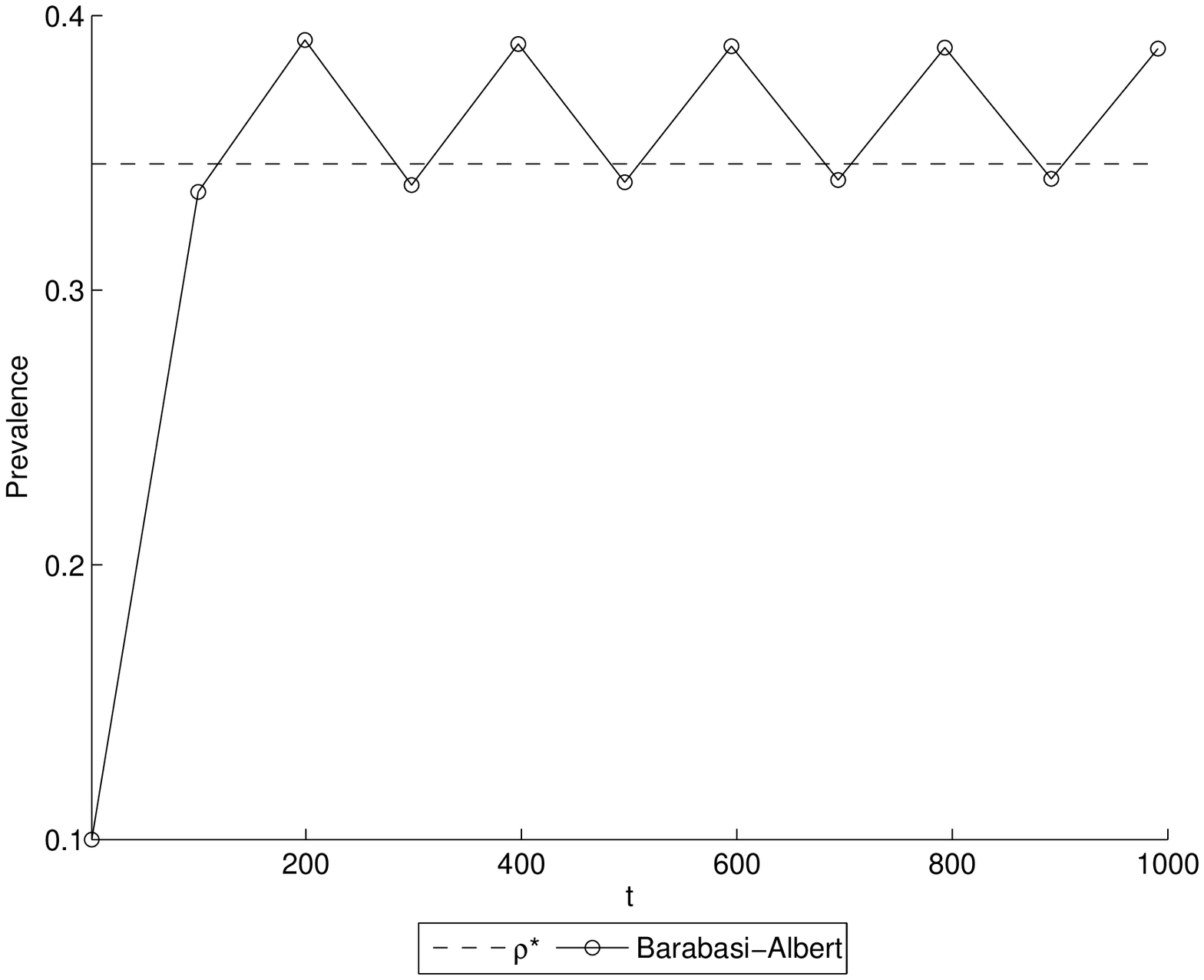}}
	\caption{\small
	Prevalence for disease transmission on BA undirected networks with 100000 vertices, $m=1$, $\lambda=0.7$ 
and $\pi_{0}=10\%$, considering a SIS model for an acute infectious disease with a fixed infectious period of one time unit.
The dashed line indicates the theoretical prediction,
	$\rho^{\star} \approx 0.346$.}
	\label{fig:figure9}
	\end{figure*}

     To compare the numerical results of the simulation with a theoretical approach, 
it is possible to deduce, for an undirected scale-free
network assembled following the BA algorithm, the equilibrium prevalence, given by~\cite{VegaRedondo}
\begin{equation}
\rho^{\star} = \frac{2}{e^{1/\lambda} - 1} \left[ 1 - \frac{1}{\lambda (e^{1/\lambda} - 1)} \right] .
\label{eq:3}
\end{equation}
This expression applies to the BA undirected network with $m=1$ and a fixed infectious period of one time unit.
For instance, for $\lambda = 0.7 $ and $\pi_{0}=10\%$, we obtain $\rho^{\star} \approx 0.346$. In Figure~\ref{fig:figure9}, 
we observe that the equilibrium prevalence in the simulation reaches the value predicted by Equation (\ref{eq:3}).

	\section{Conclusions}
	\label{sec:conclusions}
	Our approach, focusing on different networks with the same degree distribution, 
allows us to show how the topological features of a network may influence the dynamics on the network. 	
	
	Analyzing the results of the spreading simulations, we have, as expected, that the variation in the number of vertices of the hypothetical networks had little influence in the  prevalence of the diseases simulated, a result that is consistent with the characteristics of the scale-free complex networks as observed by Pastor-Satorras and Vespignani~\cite{PhysRevLett.86.3200}.
        
    With respect to the effect of the increase of the probability of spreading
on the prevalence in undirected networks (Figure~\ref{fig:figure1}), we observed that the prevalence
reaches a satured level. For undirected networks, if the probability of infection is high, 
there is a saturation of infection on the population for chronic diseases 
and therefore no new infections can occur.	

	Regarding the variation in the number of edges, the increase in the prevalence was also expected since it is known that the addition of edges increases the connectivity on the networks studied, allowing a disease to spread more easily.

	About the effect of considering the networks directed or undirected, we have that the diseases tend to stay in the undirected networks independently of the spreading model and the value of $\lambda$ considered~\cite{PhysRevLett.86.3200}, while in the directed cases due to the limitations imposed by the direction of the movements, when $m = 1$, the acute diseases tend to disappear on some of the networks.
Also due to the direction of the links, in directed networks, the disease may not reach or may even disappear
from parts of the network, 
explaining to some degree why the prevalence in directed networks (Figure~\ref{fig:figure3})
is smaller than in undirected networks (Figure~\ref{fig:figure7}).

	In the SI simulations, we could observe what would be the average maximum level of prevalence of a disease on a network and how fast this level would be reached. In the SIS simulations, the oscillations on the stability of the prevalence observed result from the simultaneous recovery of a set of vertices. With a fixed time of infection, the set of vertices that simultaneously recover is greater than in the case of a variable time of infection, since in the latter, due to the variability of the disease duration, the vertices form smaller subsets that will recover in different moments of the simulation. A result that called attention is that, in the directed networks with $m = 1$, when we simulated the chronic diseases using a fixed time, the equilibrium levels achieved were lower than the ones achieved when we used a variable time.
				
		Examining the results of the simulations on each network model, 
we have that among the undirected ones, the MB network has the lowest prevalence, with a plausible cause for this being how this network is composed, since there is a large number of vertices that are not connected to the most connected component of the network~\cite{Grisi2013}. Among the directed networks, the BA network has the lowest prevalence observed, what is also probably due to the topology of this network, since it is composed of many vertices with outgoing links only and a few vertices with many incoming and few outgoing links, thus preventing the spread of a disease.

		Using the methodology of networks, it is possible to analyze more clearly the effects that the heterogeneity in the connections between vertices have on the spread of infectious diseases, since we observed different prevalence levels in the networks generated with the same degree distribution but with different topological structures.
		
		Moreover, considering that the increase in the number of edges led to an increase in the prevalence of the diseases on the networks, we have indications that the intensification of the interaction between vertices may promote the spread of diseases. So, as expected, 
in cases of sanitary emergency, the prevention of potentially infectious contacts may contribute to control a disease.

	\section*{ACKNOWLEDGMENTS}
		This work was partially supported by FAPESP and CNPq.


{\bf Received: June 30, 2013}	
	
\end{document}